\documentclass[12pt]{iopart}
\usepackage{graphicx}

\usepackage{amstext, mathrsfs, textcomp}
\usepackage{nicefrac}
\usepackage{multirow}

\begin{document}

\title{Designing Thermoplasmonic Properties of Metallic Metasurfaces}

\author{Ch. Girard, P. R. Wiecha, A. Cuche, and E. Dujardin}
\address{CEMES, University of Toulouse and CNRS (UPR 8011), 29 rue Jeanne Marvig, BP 94347, 31055 Toulouse, France}

\date{\today}


%
\begin{abstract}
Surface plasmons have been used recently 
to generate heat nanosources, the intensity of which
can be tuned, for example, with the wavelength of the excitation radiation.
In this paper, we  
present versatile analytical and numerical investigations for the three--dimensional computation of the temperature 
rise in complex planar arrays of metallic particles.
In the particular case of elongated particles sustaining transverse and longitudinal
plasmon modes, we show a simple 
temperature rise control of the surrounding medium when turning the incident polarization.
This formalism is then used for designing novel thermoplasmonic metasurfaces for 
the nanoscale remote control of heat flux  
and temperature gradients. 
\end{abstract}
\noindent
\noindent
\maketitle
\section{Introduction}
The ability of metal structures to confine the electromagnetic fields
gave birth to a multitude of applications in areas as diverse as biophysics, 
sensor technology or devices for fast data processing\cite{Barnes:2003,Halas:2011,Brolo:2012}.
The light confinement phenomenon originates in the
surface plasmons (SP) travelling or localized at the surface of these nanostructures.
Most plasmonics applications, based on the engineering of surface plasmons, exploit
the electromagnetic fields produced by the collective electronic oscillations.
In particular, SP engineering has been considered as a viable approach to the coplanar
implementation of high speed, low dissipative information devices using analogical or digital concepts\cite{Wei:2011,Viarbitskaya:2013}.
Very recently, plasmonics has fostered another realm of applications in which dissipative
effects are being advantageously utilized
\cite{Govorov:2006,Govorov:2007,Richardson:2009,Chen:2011,Ma:2012,Baffou:2013,Cuche:2013,Coppens:2013,Levy:2014,Herzog:2014}.
Indeed, besides their widely used propensity to enhance and confine the near--field electromagnetic intensity,
metal particles and nanostructures 
have revealed a great potential as local heat sources\cite{Govorov:2006,Govorov:2007,Baffou:2013,Sanchot:2012}.
A realistic
description of such localized dissipation effects is directly related to the description of
the imaginary part of the dynamical response functions of the nanostructures, such as 
the dielectric permittivity of the metal $\boldmath{\epsilon}(\omega)$, 
and the local electric field intensity $I_{l}({\bf r},\omega)$ induced 
inside the metal\cite{Govorov:2007,Baffou:2013,Baffou:2009,Baffou:OE2009}.
While the dielectric constant only depends on the nature of the metal, 
the intensity distribution of the optical electric field induced in the particle 
is extremely sensitive to the presence of plasmon resonances occurring in the spectral
variation of the local field distribution $I_{l}({\bf r},\omega)$\cite{Govorov:2007,Baffou:2009,Girard:2008,Brosseau:2013}.
These resonances play a crucial role since 
the amount of heat tranferred to the particle can be adjusted by tuning
the incident wavelength in or out of the resonance range.
Several experimental thermoplasmonic building blocks
have indeed been designed from this concept and realized from colloidal chemistry or sophisticated
lithography processes
\cite{Coppens:2013,Levy:2014,Sanchot:2012,Baffou:2013a}. 

In this paper, we propose a flexible analytical scheme \cite{Teulle:2013,Weber:2015} 
completed by 
numerical studies \cite{Girard:2008} 
to investigate the thermoplasmonic properties  of arrays of individual 
plasmonic entities with arbitrary shapes sustaining multiple plasmon
modes in the optical range.
Arrays of nanostructures are systems of fundamental interest in plasmonics
since they combine the optical properties of individual resonators
and the collective response of the assembly. 
Such systems have already contributed to major breakthroughs in several fields in optics
like optical sensing, metasurfaces for light phase and orbital angular momentum control, strong optical coupling, ...
\cite{Enkrich:2005,Yu:2011,Lawrence:2012,Anker:2008}.
In a typical configuration,
the particle arrangement is supported by a planar dielectric substrate. Here 
we apply the well--established self--consistent scheme based on the
Green Dyadic Functions ({\sf GDF}) formalism and compute the local field
intensity inside the particle array by including the coupling with the substrate. 
The average temperature in the vicinity of the metallic sructures
is then derived from the local field intensity. 

As a first step towards realistic configurations, we describe each 
individual metal structure as an anisotropic polarizable particle excited by 
their self--consistent local electric field\cite{Book-CG-ED:2010}. 
This first approach consists in gradually developping a quasi--analytical description
of the calculation that provides an intuitive access to the underlying physical 
and thermal mechanisms. 
This simple analytical description is then complemented by an adequate discretisation of the particle physical volumes
in order to describe arbitrary geometries\cite{Girard:2008}. 
The numerical applications are based on the permittivity
of gold taken from Johnson and Christy data\cite{Johnson:1972}.
In the third section, the specific case of periodic arrays of gold nanorods is investigated and we
demonstrate a simple and reversible control of the temperature rise 
near the particles
when turning the incident polarization.  
Applications to the new concept of thermoplasmonic metasurfaces are then discussed in the two last sections
where we demonstrate that our numerical
technique is well--suited for the design of optimized thermoplasmonic meta--cells,
using an evolutionary optimization ({\sf EO}) algorithm.
\begin{figure}[h!]
\centering\includegraphics[width=10cm]{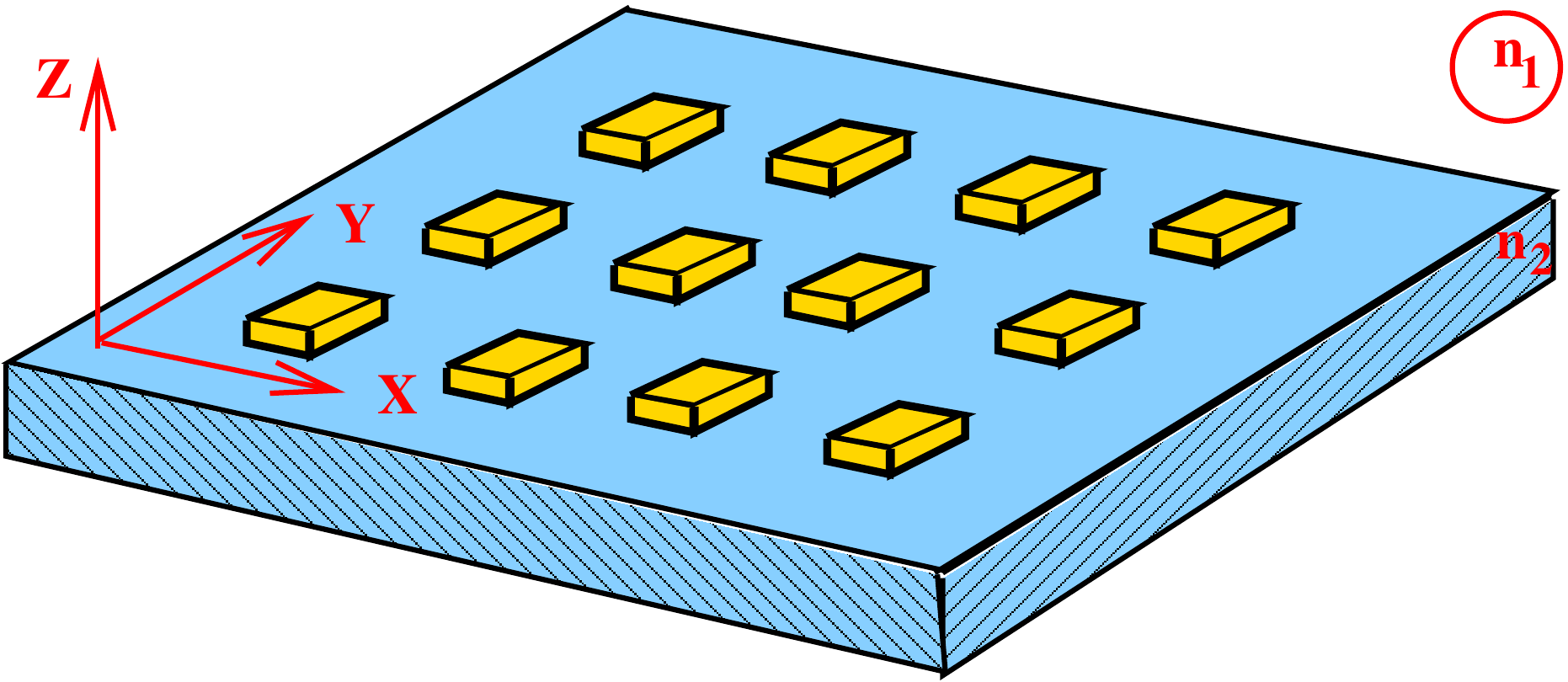}
\caption{(color online) Perspective view illustrating a periodic assembly
of plasmonic structures fabricated at the surface of an insulating
sample.
}
\label{FIG1}
\end{figure}
\section{Thermal response of a periodic array of identical metal particles} 
Let us consider a periodic {\sf 2D} array 
of $N$ elongated
gold structures arranged in a periodic way at the surface of a dielectric planar substrate (Fig. (\ref{FIG1})).
The particle location is defined by a set of  N vectors ${\bf r}_{i}$ = $({\bf L}_{i},Z)$
(where the two--dimensional vector ${\bf L}_{i}$ belongs to the $(XOY)$ plane).
Unlike what happens with perfectly spherical particles,
single nanorods exhibit extinction spectra with two plasmon bands
that correspond to electron oscillations along their length
(low energy longitudinal mode) and across their section (high energy transverse mode) \cite{Link:1999}.
This shape effect can be described with a simple analytical model
by using an anisotropic dynamical polarisability \cite{Book-CG-ED:2010}.
When the long axis  of the particles is aligned along (OY) axis 
as shown in figure (\ref{FIG1}), the polarizability tensor
is diagonal and reads:
\begin{eqnarray}
\alpha(\omega_{0})=
\left (
\begin{array}{ccc}
\alpha_{\perp}(\omega_{0}) & 0 & 0\\
0 &  \alpha_{\parallel}(\omega_{0})& 0\\
0 & 0 &\alpha_{\perp}(\omega_{0})
\end{array}
\right)
\; ,
\end{eqnarray}
where the two independent components $\alpha_{\perp}(\omega_{0})$ and
$\alpha_{\parallel}(\omega_{0})$ can be described by the formula associated
with a prolate ellipsoid \cite{Book-CG-ED:2010}.
\subsection{Local field calculation}
When a monochromatic electromagnetic plane wave of frequency 
$\omega_{0}$  and electric field amplitude ${\bf E}_{0}$ 
hits the interface between environment (media 1) and the glass substrate (media 2) 
at normal incidence and interacts with the metallic particles, the optical electric field can be written:
\begin{equation}
{\bf E}_{0}({\bf r},t)=\frac{1}{2}\{{\bf E}_{0}({\bf r},\omega_{0})\exp (i\omega_{0}t) +C.C.\}
\; ,
\label{FIRST-E0}
\end{equation}
where ${\bf E}_{0}({\bf r},\omega_{0})$ (with ${\bf r}$ = $(x,y,z)$) represents its Fourier amplitude:
\begin{equation}
{\bf E}_{0}({\bf r},\omega_{0})={\bf E}_{0}[\exp(-in_{1}k_{0}z)+R\exp(in_{1}k_{0}z)]
\; ,
\label{EFA}
\end{equation}
in which $k_{0}$ is the wave vector modulus in vacuum and $R$
= $(n_{1}-n_{2})/(n_{1}+n_{2})$
is the Fresnel reflection coefficient expressed with the optical indices 
of surrounding medium ($n_{1}$) and dielectric substrate ($n_{2}$), respectively.
The polarization of the incident wave, associated with the direction of the vector ${\bf E}_{0}$, can be materialized by
the angle $\theta$ between ${\bf E}_{0}$ and the (OX) axis:
\begin{equation}
{\bf E}_{0}=E_{0}(\cos(\theta),\sin(\theta),0)
\; .
\label{ANG}
\end{equation}
\subsubsection{Self--consistency}
The local fields ${\bf E}({\bf r}_{i},\omega_{0})$ induced
at the center of the particles by the illumination field verify a set of $N$ coupled equations that can be condensed as follows:
\begin{equation}
{\cal E}_{0}(\omega_{0})={\cal M}(\omega_{0})\cdot {\cal E}(\omega_{0})
\label{EQ-SELF}
\end{equation}
where  ${\cal E}_{0}(\omega_{0})$ is the {\it input} supervector that contains the $N$ incident fields at
the particle locations:
\begin{equation}
{\cal E}_{0}(\omega_{0})=
\{{\bf E}_{0}({\bf r}_{1},\omega_{0}),....,{\bf E}_{0}({\bf r}_{i},\omega_{0}),...\}
\; ,
\end{equation}
and ${\cal E}(\omega_{0})$ is the {\it output} supervector that contains the local field values:
\begin{equation}
{\cal E}(\omega_{0})=
\{{\bf E}({\bf r}_{1},\omega_{0}),....,{\bf E}({\bf r}_{i},\omega_{0}),...\}
\; .
\label{LOCAL-F}
\end{equation}
For $N$ polarizable particles, the (3$N$ $\times$ 3$N$) coupling matrix ${\cal M}(\omega_{0})$ has a very simple form given by:
\begin{equation}
{\cal M}(\omega_{0})={\cal I}-{\cal A}(\omega_{0})
\; ,
\end{equation}
where ${\cal I}$ is the identity matrix and:
\begin{eqnarray}
{\cal A}(\omega_{0})=
\left (
\begin{array}{cccc}
{\bf A}_{11}(\omega_{0}) & .... & {\bf A}_{1j}(\omega_{0})  & ....\\
.... & .... & .... & ....\\
.... & {\bf A}_{ij}(\omega_{0}) & .... & ....\\
.... & .... & .... & {\bf A}_{NN}(\omega_{0}) 
\end{array}
\right)
\; ,
\end{eqnarray}
is composed of $N^{2}$ submatrices defined from the particle polarizabilities
and the field--propagators ${\bf S}({\bf r}_{i},{\bf r}_{j},\omega_{0})$ 
between two particles locations, ${\bf r}_{i}$ and ${\bf r}_{j}$:
\begin{equation}
{\bf A}_{ij}(\omega_{0})={\bf S}({\bf r}_{i},{\bf r}_{j},\omega_{0})\cdot\alpha(\omega_{0})
\end{equation}
\subsubsection{Weak coupling between individual metallic nanotructures}
The calculation of the local field in the particle array needs
the inversion of the matrix ${\cal M}(\omega_{0})$ shown by equation (\ref{EQ-SELF}),
which can be considerably simplified depending on the interparticle spacing $D$.
For example, when the lateral spacing $D$ is of the order of the incident wavelength $\lambda_{0}$
and the thickness of the metal particles is much smaller,
the mutual interactions vanish so that 
all the tensorial components ${\bf S}({\bf r}_{i},{\bf r}_{j},\omega_{0})\cdot\alpha(\omega_{0})$
$\ll$ 1. This hypothesis, that corresponds to the first Born approximation
({\sf FBA}), leads to the simplified relation: 
\begin{equation}
{\cal M}^{-1}(\omega_{0})={\cal I}+{\cal A}(\omega_{0})+
O({\cal A}^{2})
\; ,
\end{equation} 
\subsection{Dissipated power}
During the illumination process, the temperature 
rises because of the electronic Joule effect induced inside the metal particles.
This dissipative energy channel can be described by
computing the power per unit volume dissipated
inside the metal. 
From the electric field ${\bf E}_{i}({\bf r},t)$ 
and the induction vector ${\bf D}_{i}({\bf r},t)$,  
we can derive the amount of power
dissipated by the i$^{th}$ metallic particle. In {\sf CGS} electrostatic
units, this leads to:
\begin{equation}
{\cal Q}({\bf r}_{i})=\frac{1}{4\pi}\int_{v_{i}}d{\bf r}
<{\bf E}_{i}({\bf r},t)\cdot\frac{\partial}{\partial t}{\bf D}_{i}({\bf r},t)>
\; ,
\label{Q1}
\end{equation}
where the integral runs over the particle volume, and the brakets
schematize the time average. 
Similarly to equation (\ref{FIRST-E0}),
the vectors ${\bf E}_{i}({\bf r},t)$ and ${\bf D}_{i}({\bf r},t)$ 
can be expressed in term of their 
Fourier amplitudes ${\bf E}_{i}({\bf r},\omega_{0})$ and ${\bf D}_{i}({\bf r},\omega_{0})$. 
After taking the time average, this transformation leads to:
\begin{eqnarray}
\label{Q2}
{\cal Q}({\bf r}_{i})=\frac{1}{16\pi}\int_{v_{i}}
i\omega_{0}\Big{\{} {\bf E}_{i}({\bf r},\omega_{0})\cdot{\bf D}_{i}^{\star}({\bf r},\omega_{0})
\\ \nonumber
-{\bf E}^{\star}_{i}({\bf r},\omega_{0})\cdot{\bf D}_{i}({\bf r},\omega_{0})\Big{\}}d{\bf r}
\; ,
\end{eqnarray}
Next, we introduce the constitutive
relation between ${\bf D}_{i}({\bf r},\omega_{0})$ and ${\bf E}_{i}({\bf r},\omega_{0})$. 
Rewriting this
equation in a tensorial form will allow to consider non--spherical particles:
\begin{equation}
{\bf D}_{i,\alpha}({\bf r},\omega_{0})=
\sum_{\beta}{\epsilon}_{\alpha,\beta}(\omega_{0}){\bf E}_{i,\beta}({\bf r},\omega_{0})
\; ,
\label{EQ-CONS}
\end{equation}
where $\alpha$ and $\beta$ are two cartesian indices.
After replacing (\ref{EQ-CONS}) into (\ref{Q2}) and assuming a diagonal form
for ${\epsilon}_{\alpha,\beta}(\omega_{0})$ one gets:
\begin{equation}
{\cal Q}({\bf r}_{i})=\frac{\omega_{0}}{8\pi}\int_{v_{i}}
\Big{\{} \sum_{\alpha} \Im {\epsilon}_{\alpha,\alpha}(\omega_{0})|{\bf E}_{i,\beta}({\bf r},\omega_{0})|^{2} 
\Big{\}} d{\bf r}
\; ,
\label{Q3}
\end{equation}
where the $\Im$ symbol means {\it imaginary part}.
Equation (\ref{Q3}) is general and does not contain any approximation and applies to arbitrary
geometries and any type of materials.
\subsubsection{Dipolar response  approximation}
For metallic particles of small size, the multipolar contributions higher 
than the dipolar one can be neglected.
In this case, the permittivity of the i$^{th}$  particle can be schematized by the following relation
in which $\delta({\bf r}-{\bf r}_{i})$ represents the Dirac $\delta$ distribution centered
around the particle location ${\bf r}_{i}$:
\begin{equation}
{\epsilon}(\omega_{0})=1+4\pi 
\left (
\begin{array}{ccc}
\alpha_{\perp}(\omega_{0}) & 0 & 0\\
0 &  \alpha_{\parallel}(\omega_{0})& 0\\
0 & 0 &\alpha_{\perp}(\omega_{0})
\end{array}
\right)
\times \delta({\bf r}-{\bf r}_{i}) 
\; .
\label{EPS-EXP}
\end{equation}
From this simplified relation it is straightforward to perform the volume integral
found in equation (\ref{Q3}):
\begin{eqnarray}
{\cal Q}({\bf r}_{i})=\frac{\omega_{0}}{2}
\Big{\{} \Im \alpha_{\perp}(\omega_{0}) |{\bf E}_{i,y}({\bf r}_{i},\omega_{0})|^{2}
\nonumber
\\
+ \Im \alpha_{\parallel}(\omega_{0}) (|{\bf E}_{i,x}({\bf r}_{i},\omega_{0})|^{2}
+ |{\bf E}_{i,z}({\bf r}_{i},\omega_{0})|^{2}) \Big{\}} 
\; ,
\label{Q4}
\end{eqnarray}
Notably, a more accurate calculation of both the local fields and 
the successive field–-gradients would require to go beyond the dipolar approximation. 
However, the proposed description makes it possible 
to derive the analytical calculation throughout its development.
\subsubsection{The particular case of a single elongated plasmonic particle}
When the metallic pattern simply consists in a single particle located at the position
${\bf r}_{1}$, the collective effects vanish and
the matrix ${\cal M}^{-1}(\omega_{0})={\cal I}$ which means that the local electric field 
${\bf E}({\bf r}_{1},\omega_{0})$ = ${\bf E}_{0}({\bf r}_{1},\omega_{0})$. 
This asymptotic case gives rise to a particularly simple equation that can be obtained
by using Eqs. (\ref{EFA}) and (\ref{ANG}):
\begin{equation}
{\cal Q}=\frac{E_{0}^{2}\omega_{0}}{2}
\{\Im \alpha_{\perp}(\omega_{0}) \cos^{2}(\theta)
+\Im \alpha_{\parallel}(\omega_{0}) \sin^{2}(\theta) \} 
\label{Q-SING}
\; .
\end{equation}
Here, we explicitly see how tuning the polarization angle $\theta$
controls the amount of heat transferred to the particle. 
Finally, after introducing the well--known relation between electric field amplitude
$E_{0}$ and laser power $S_{0}$ delivered per unit area \cite{Landau:1960a}:
\begin{equation}
E_{0}^{2}=\frac{8\pi S_{0}}{c}
\; ,
\label{S0}
\end{equation}
one obtains:
\begin{equation}
{\cal Q}=4\pi S_{0}k_{0}
\{\Im \alpha_{\perp}(\omega_{0}) \cos^{2}(\theta)
+\Im \alpha_{\parallel}(\omega_{0}) \sin^{2}(\theta) \} 
\label{Q-SING1}
\; .
\end{equation}
In this last equation, the influence of the relative weights of both transverse 
and longitudinal plasmon modes clearly appears through the two components of the nanorod polarizability.
\begin{figure}[h!]
\centering\includegraphics[width=10cm]{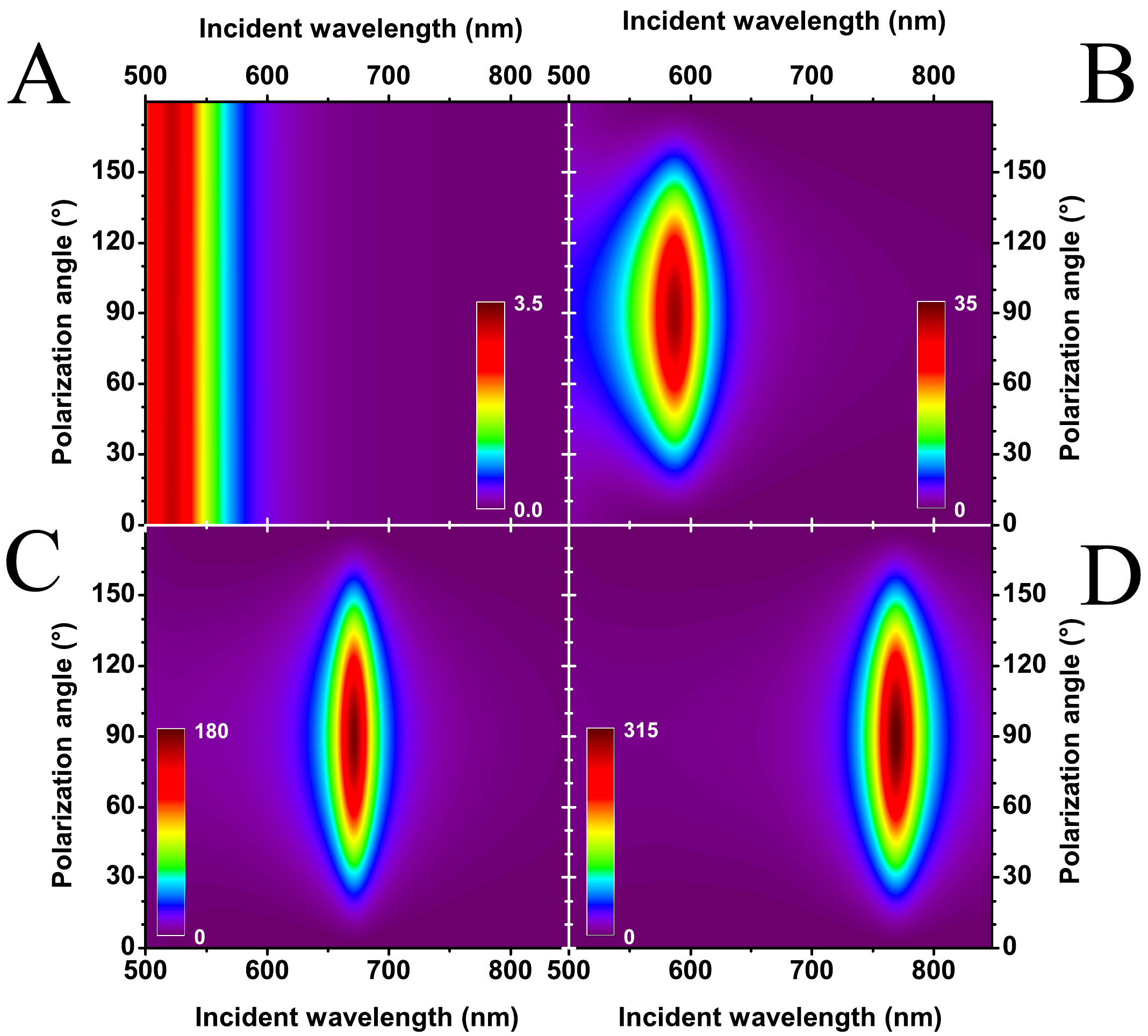}
\caption{(color online) 
Simulation of four color plots of the heat ${\cal Q}(\lambda_{0},\theta)$ dissipated per time unit
as a function of the incident wavelength (500 nm $\leq$ $\lambda_{0}$ $\leq$ 850 nm)  and the polarization angle
(0 $\leq$ $\theta$ $\leq$ 180$^{o}$) for a gold prolate ellipsoid of short and long axis $b$ and $a$,
respectively. In maps (A) to (D), four aspect ratio $\eta$ = $a/b$ have been considered: 
(A) sphere case ($\eta$ =1)  $a$ = $b$ = 10 nm; (B) ellipsoid ($\eta$ =2) $a$ = 20 nm and $b$ = 10 nm;
(C) ellipsoid ($\eta$ =3) $a$ = 30 nm and $b$ = 10 nm; and (D) ellipsoid ($\eta$ =4) $a$ = 40 nm and $b$ = 10 nm.
All the color bars are expressed in nanoWatt (nW).
}
\label{FIG2}
\end{figure}
To illustrate how this equation governs 
the heat absorbed by a single ellipsoidal gold nanorod, we present in Fig (\ref{FIG2})
a sequence of four maps ${\cal Q}(\lambda_{0},\theta)$ for aspect ratios $\eta$ = $a/b$
varying from 1 to 4. The laser power $S_{0}$ is set at 1 $mW/\mu m^{2}$
and the particle long axis is chosen parallel to the $(OY)$ cartesian axis. 
In the ($\lambda_{0},\theta$) coordinate plane, the resulting heat response of the particle 
varies from 
polarization--independent dissipation maximized 
at low incident wavelength for the sphere case (characterized by
a single color band), to an oblong domain that is 
shifted to longer incident wavelength as $\eta$ increases and shows an optimum for 90$^{o}$ polarization angle.
\subsubsection{Volume discretization approach}
Finding exact solutions of equation (\ref{Q1}) for more realistic situations
requires an additional volume discretization procedure of the source region occupied by the plasmonic
particles. Generally, each particle volume $V_p$ is discretized into $n_p$ identical elementary volumes $v_p$.
Such a procedure converts integrals into discrete summations.
The main analytical steps of this technique are detailed in reference\cite{Girard:2008} and lead to:
\begin{eqnarray}
\label{DISCRE}
{\bf E}({\bf r}_{t,i},\omega_{0})={\bf E}_{0}({\bf r},\omega_{0})
+\sum_{p=1}^{N}\chi_{p}(\omega_{0})\sum_{j=1}^{n_{p}}
\\
\nonumber
\times {\bf S}({\bf r}_{t,i},{\bf r}_{p,j},\omega_{0})
\cdot {\bf E}({\bf r}_{p,j},\omega_{0})
\; .
\end{eqnarray}
In this expression, the parameters $\chi_{p}(\omega_{0})$ associated with the elementary volumes $v_p$ 
are homegeneous to  dipolar polarizabilities:
\begin{equation}
\chi_{p}(\omega_{0})=\frac{\epsilon_{p}(\omega_{0})-\epsilon_{env}(\omega_{0})}{4\pi}v_{p}
\; .
\label{ETAPOL}
\end{equation}
The vectors ${\bf r}_{p,j}$ and ${\bf r}_{t,i}$ represent the location of $j^{th}$ and $i^{th}$ discretized cells
inside the $p^{th}$ and $t^{th}$ metallic particles, respectively.
Next, the self--consistent electric field inside the metal particles is computed. This procedure
leads to a system of $N\times n_{p}$ vectorial equations with $N\times n_{p}$ unknown fields
${\bf E}({\bf r}_{p,j},\omega_{0})$.

For a given metal particle (labelled by the subscript $p$),
the solving procedure  detailed in reference\cite{Girard:2008} 
is directly related to the discretization volume
$v_{p}$, which itself depends on the discretization grid used to mesh the particles. 
The expressions of the $\chi_{p}(\omega_{0})$ coefficients for both cubic and hexagonal 
compact discretization grids can be found in table (1) of reference \cite{Girard:2008}.
\section{Temperature Profile}
\begin{figure}[h!]
\centering\includegraphics[width=10cm]{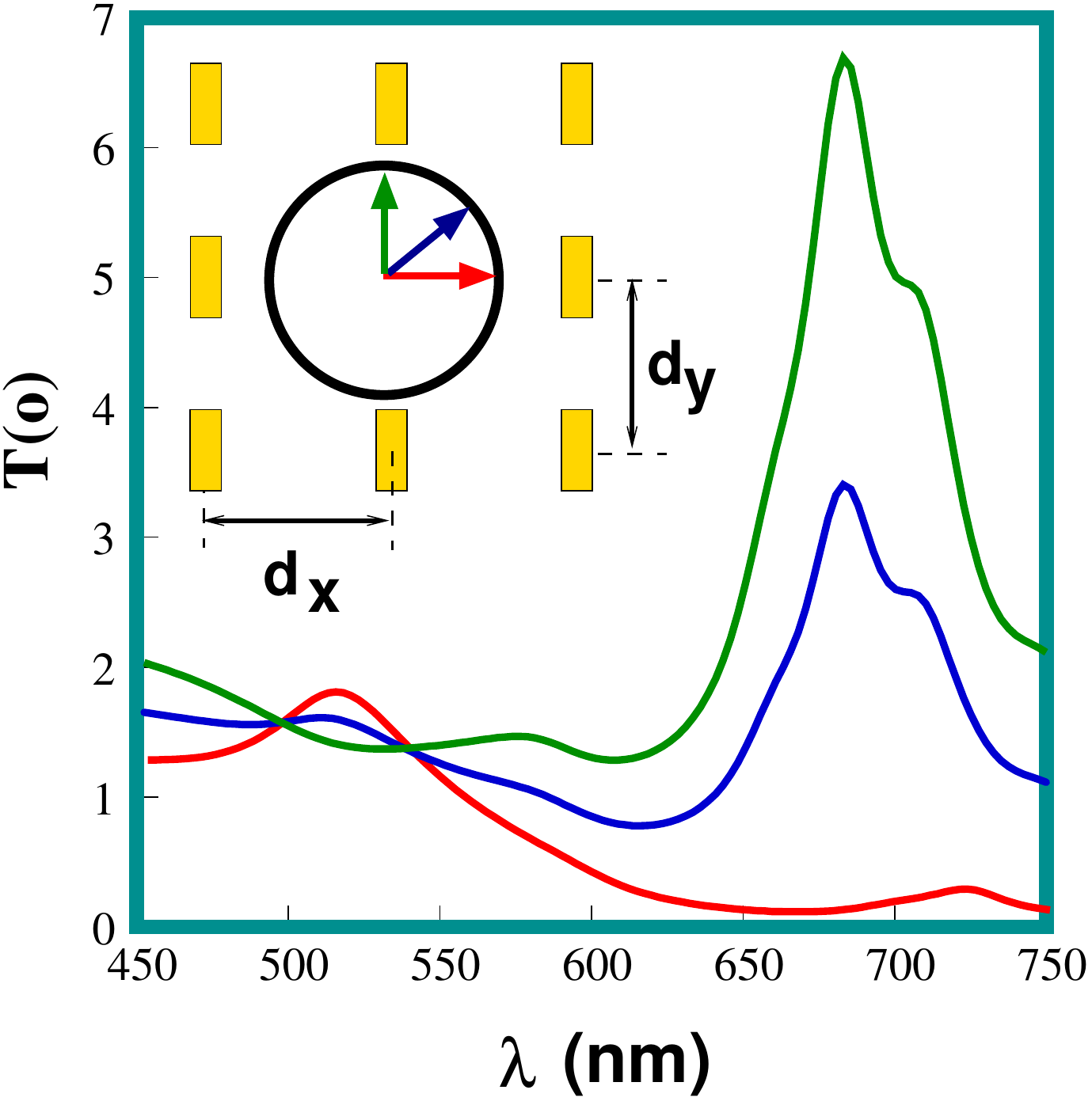}
\caption{(color online)
Simulation of three  temperature spectra 
computed above a set of nine cylindric gold nanorods (25 $\times$ 80 nm)
illuminated in normal incidence by a linearly polarized plane wave
(see geometry insert). The laser power $S_{0}$ is fixed at 5 $mW/\mu m^{2}$
and the lateral pitches, $d_{x}$ and $d_{y}$ between the rods 
is 500 nm. 
The three spectra correspond to the polarization directions indicated by the red, blue and green arrows.
The computation has been performed by discretization of the nanorod volumes
by N $\times$ 333 elementary cells distributed over a hexagonal compact three--dimensional mesh. 
}
\label{FIG3}
\end{figure}
\begin{figure}[h!]
\centering\includegraphics[width=10cm]{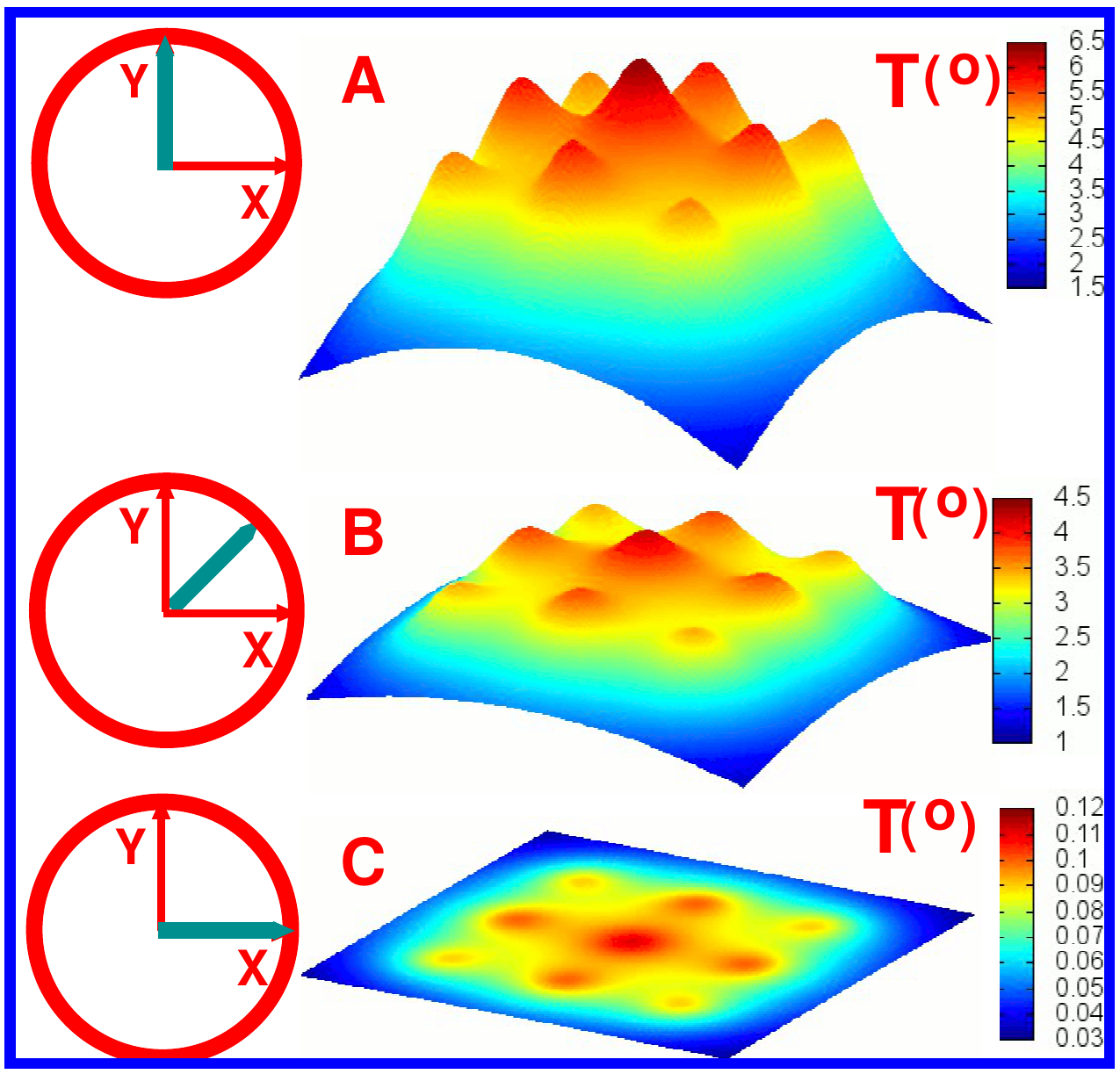}
\caption{(color online)
3D color plots of three simulations of the temperature 
distribution above a set of nine gold nanorods corresponding to incident polarization aligned (A) along Y axis, 
(B) at 45$^{o}$ and (C) along the X axis (same parameters as figure (\ref{FIG3})).
The wavelength, $\lambda_{0}$ = 680 nm, is chosen at the center of the longitudinal plasmon band. 
}
\label{FIG4}
\end{figure}
The steady–-state temperature increase, $\Delta T(\lambda_{0},\theta)$,  
can be deduced from the distribution of heat ${\cal Q}({\bf r}_{p,j})$
deposited inside the particle lattice (cf. Eq. (\ref{Q3})), by the thermal Poisson equation:
\begin{equation}
\Delta T(\lambda_{0},\theta)=\frac{1}{4\pi \kappa}
\sum_{p=1}^{N}\sum_{j=1}^{n_{p}}\frac{{\cal Q}({\bf r}_{p,j})}{|{\bf R}_{obs}-{\bf r}_{p,j}|}
\label{TEMP}
\end{equation}
where $\kappa$ is the environmental thermal conductivity and
${\bf R}_{obs}$ defines an observation point located in the vicinity
of the sample. In figures (\ref{FIG3}) and (\ref{FIG4}), we have used 
relation (\ref{TEMP}) to investigate the photothermal effects
induced near an array of nine gold nanorods deposited on a dielectric surface
(see insert of figure (\ref{FIG3})). The metal particles are surrounded by
an isotropic medium of refractive index $n_{1}$ = 1.33,
mimicking an aqueous medium. 

The spectral variation of the temperature as a function of the incident wavelength
is presented in figure (\ref{FIG3}).
The temperature shift $\Delta T(\lambda_{0},\theta)$ is computed from equation (\ref{TEMP})
at a position ${\bf R}_{obs}$ = $(0,0,150 nm)$ which overhangs the central gold pad.
The three polarisation directions considered here, {\it i.e.} zero degree (red curve); 45$^{o}$ (blue curve)
and 90$^{o}$ (green curve), demonstrate that the temperature 
around the metal pattern can
be effectively tuned with a drastic increase observed near the longitudinal resonance. 
Thus, by exciting the longitudinal plasmon band ($\lambda$ = 680 nm)
of the sample, the local temperature can be modulated over one order of magnitude by the simple tuning of the the field polarization. 
Obviously, this control is much less effective when exciting the transverse mode ($\lambda$ = 530 nm)
because of a weaker quality factor. In addition this 
resonance is bound by two isobestic points \cite{Baffou-acs-2017}  at 500 and 540 nm, where the temperature 
is independent of the polarization (Figure \ref{FIG3}).

Finally, figure (\ref{FIG4}) shows a sequence of three temperature maps resulting from the
monochromatic excitation of the longitudinal band
($\lambda$ = 680 nm) at normal incidence with a plane wave. 
All three maps are displayed with the same vertical 
scale to highlight the temperature rise occurring when the incident 
polarization is aligned with the main axis of the nanorods. 
A general observation is the strong temperature rise in the direct vicinity of individual metallic 
structures and the broader yet less intense temperature increase over the entire pattern \cite{Baffou:2013a}. 

\section{Metasurfaces for Thermoplasmonic control}
\label{META}
\begin{figure}[h!]
\centering\includegraphics[width=10cm]{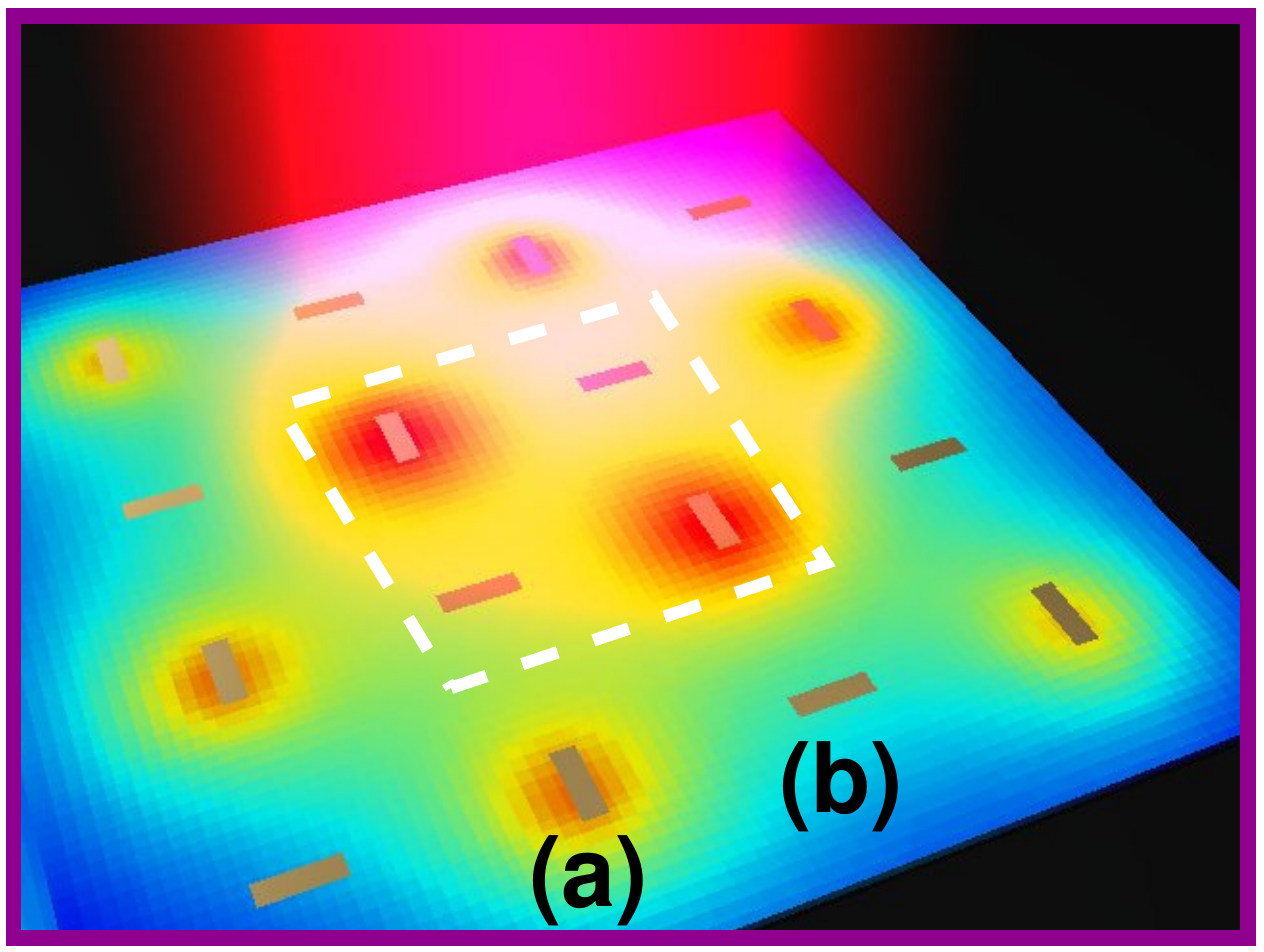}
\caption{(color online)
Example of thermoplasmonic metasurface able to generate
strong temperature contrast. 
The unit cell, located inside the dashed white frame, is a set of four gold nanorods perpendicular to each other.
The labels (a) and (b) represents two consecutive rod orientations, parallel and antiparallel, inside the pattern.
The metasurface is superimposed by a temperature map computed with a incident polarization aligned 
along the (a) nanorods. 
}
\label{FIG5}
\end{figure}
In the recent years, by designing the surface of some materials at a subwavelength scale,
new applications for optics were highlighted (see for example \cite{Capasso:2014}).
These new structures have been referred to as metasurfaces because they 
can modify the main physical characteristics
of the incident light. These modifications include, the phase, the angular momentum, or the light polarization
\cite{Black:2014}, and can generate and exalt nonlinear phenomena \cite{Linden:2015,Minovitch:2015,Black:2015}.
For example, plasmonic metasurfaces containing two--dimensional subwavelength gold patterns
have been developed that allow imprinting arbitrary phase patterns onto a propagating beam \cite{Yu:2011}
and perpendicular gold nanorods arrays have been used to perform polarization conversion
control from
the capacitive coupling to the conductive coupling regimes between the gold entities \cite{Black:2014}.

Metasurface physics also presents attractive opportunities for the thermoplasmonics.
In particular, the design and the juxtaposition of elementary plasmonic cells sensitive to the incident polarization is a
good manner to lead to efficient temperature gradient controls.
In figure (\ref{FIG5}), we consider a simple example of metasurface consisting of elementary cells containing four nanorods.
In this paving, two consecutive plasmonic elements are perpendicular relative to each other, 
so that the excitation of the transverse and longitudinal modes will move 
from one structure to its neighbor, when progressively turning the polarization angle of the incident light.
The next figure (\ref{FIG6}) displays the temperature map evolution expected around this  metasurface geometry. 
These computations have been performed by keeping the
same parameters as in the previous section, i.e.,   
16  cylindric gold nanorods (25 $\times$ 80 nm)
illuminated in normal incidence by a linearly polarized plane wave
(see geometry insert). The laser power $S_{0}$ is fixed at 5 $mW/\mu m^{2}$
and the lateral pitches, $d_{x}$ and $d_{y}$ between the rods
are 250 nm.
A detailed examination of the maps shows a regular shaping of 
the temperature distribution, in a range that varies from 5 degrees for an incident power
of 5 $mW/\mu m^{2}$, and that displays a periodic series of hot spots
with tunable location 
by applying a remote control of the polarization.
For example, as shown in the first map of figure (\ref{FIG6}), a polarization parallel to $OX$ axis
yields a hot spot pattern distributed on a square lattice with a side  equal to 
$\sqrt{2}$ $\times$ the lateral nanorod pitch $d_{x}$.
Consequently, by gradually turning the incident polarization, we can accurately
drag these hot spots, in a controllable manner, from the parallel to perpendicularly oriented nanorods.
In addition, as described in the {\it Supplemental Information Document} (see figures S2 and S3),
such a local temperature gradient tuning leads to the possibility of controlling
the heat flux in the vicinity of the metasurface. 

This functionality is particularly attractive for nano–-biology manipulations and applications, 
where tuning the symmetry of such temperature gradients 
would generate complex convection currents in liquids that could be advantageously exploited, 
for example, to thermally assisted plasmonic trapping \cite{Donner:2015,Cuche-prl:2012}, 
to trigger thermotactic mobility or behavioral plasticity on demand \cite{Mori:1995,Hamada:2008}.

\begin{figure}[h!]
\centering\includegraphics[width=10cm]{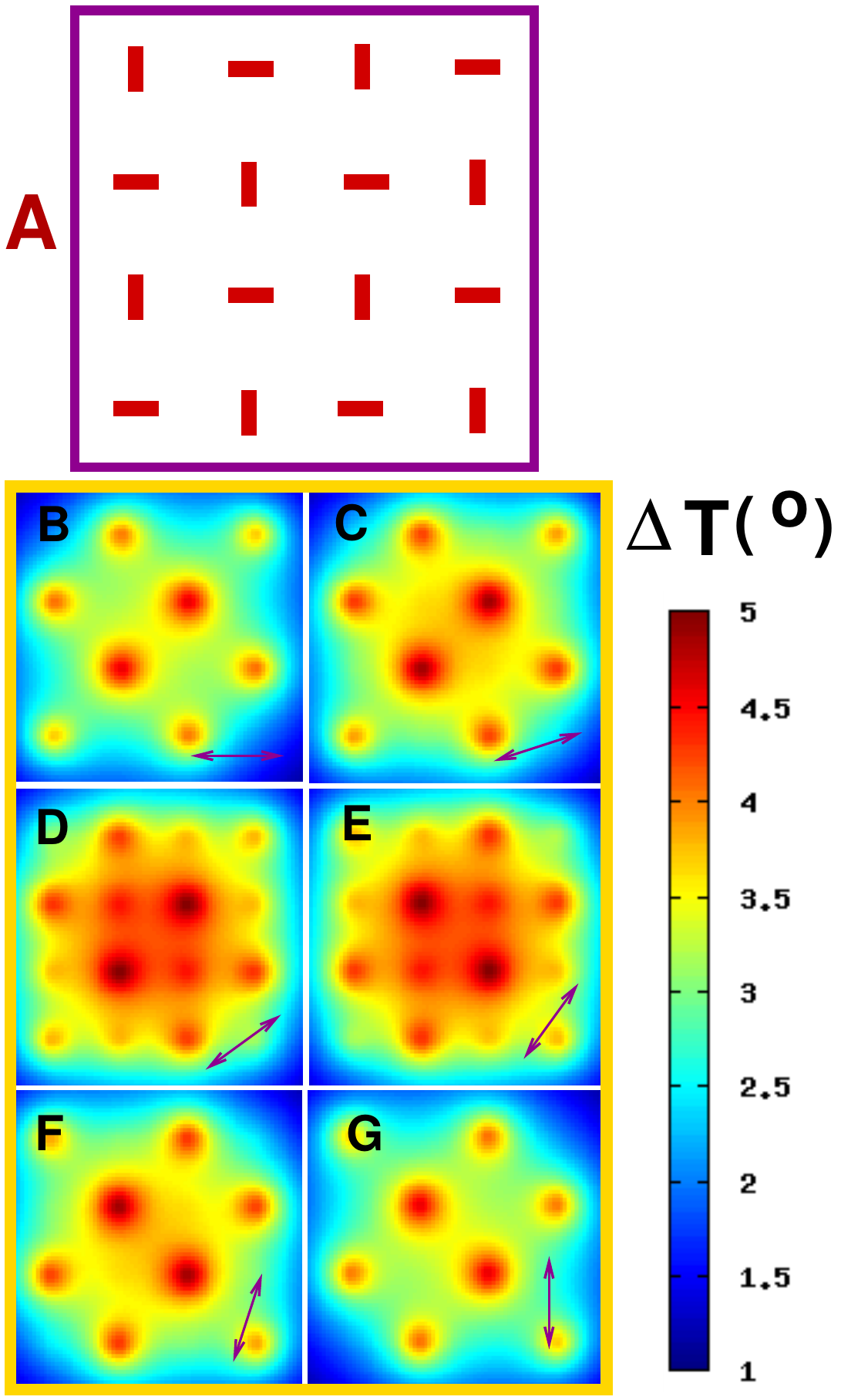}
\caption{(color online)
(A) Top view of the thermoplasmonic metasurface described in figure (\ref{FIG5}).  
(B) to (G) evolution of the temperature maps when the incident polarization,
represented by a double red arrow, 
is turned from $0^{\deg}$ to $90^{\deg}$ with  $18^{\deg}$ steps. 
}
\label{FIG6}
\end{figure}
\section{Evolutionary optimization of optical hybrid material meta--cells}
In order to complete this theoretical paper, we demonstrate that our numerical 
technique is well--suited for the design of optimized thermoplasmonic meta--cells, 
using an evolutionary optimization ({\sf EO}) algorithm.
In the recent past, {\sf EO} techniques have been successfully 
applied on various problems in nano--optics\cite{feichtner_evolutionary_2012,forestiere_inverse_2016,wiecha_evolutionary_2017}.
We will also use a multi-material structure in our demonstration, 
which means that each meta-cell (or meta-unit) is composed of multiple elements of different materials.
Multi-material systems can be modeled with our approach 
by using position-dependent 
dielectric functions $\epsilon_p(\omega_0)$ in equation (\ref{ETAPOL}) 
for the meshpoints at $\mathbf{r}_p$. 
A full metasurface would finally consist of many of those optimized meta-units.
\subsection{Evolutionary optimization of photonic nano-structures}
As illustrated in our previous examples (see section (\ref{META})),
the design of geometric assemblies of nanostructures starts with the conception of a reference geometry
by intuitive considerations.
Via the systematic variation of a few parameters, this reference systems is then optimized within its possibilities.
Such an approach, however, is limited to rather simple problems and requires a certain degree of understanding and intuition
for the considered system.
In case of complex structures or complicated phenomena, the intuitive method often fails.
To overcome these limitations, we apply in this section
an evolutionary optimization algorithm in order to design a meta--unit for optimum
nano--scale heat generation.
{\sf EO} is a heuristic optimization method to find the global maximum or minimum of complex, possibly non-analytic problems.
{\sf EO} algorithms use a \textit{population} of parameter--sets for the problem, which are driven through 
a cycle of reproduction, evaluation and selection, 
in which weak solutions are iteratively eliminated and {\it strong} parameter-sets are kept.
Here, we couple the ``jDe'' {\sf EO} algorithm 
\cite{islam_adaptive_2012} provided by the ``paGMO'' toolkit 
\cite{biscani_global_2010} to our volume discretization method for full--field electrodynamical simulations.
For these simulations, we use our own toolkit ``pyGDM'' \cite{wiecha_pygdm_2018}.
More details on the approach can be found in Ref.~\cite{wiecha_evolutionary_2017}.
\begin{figure}[h!]
\centering\includegraphics[width=14cm]{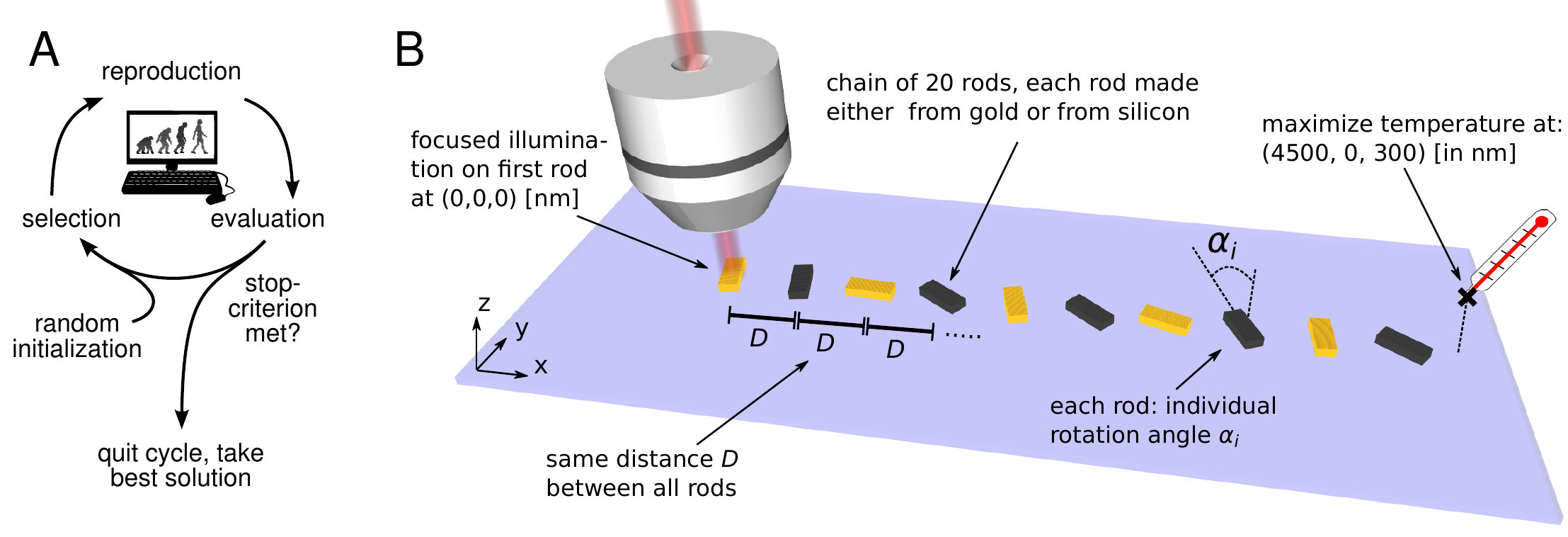}
\caption{(color online)
A) sketch of the evolutionary optimization scheme.
(B) illustration of the optimization model and problem.
A chain of $20$ nano-rods on a glass substrate is searched in order 
to maximize the temperature increase at a specific location 
$(x=4500,y=0,z=300)$ [nm], far from the focal spot of a focused illumination (at $(x=0,y=0,z=0)$ [nm]).
The free parameters of the optimization 
are the angles $\alpha_i$ as well as the material (either gold or silicon) of each rod. 
As further parameter, the spacing $D$ between the rods is optimized by the algorithm.
}
\label{fig:sketch_opt}
\end{figure}
\subsection{Optimization of a chain of nano-rods for localized heating}
\begin{figure}[h!]
\centering\includegraphics[width=14cm]{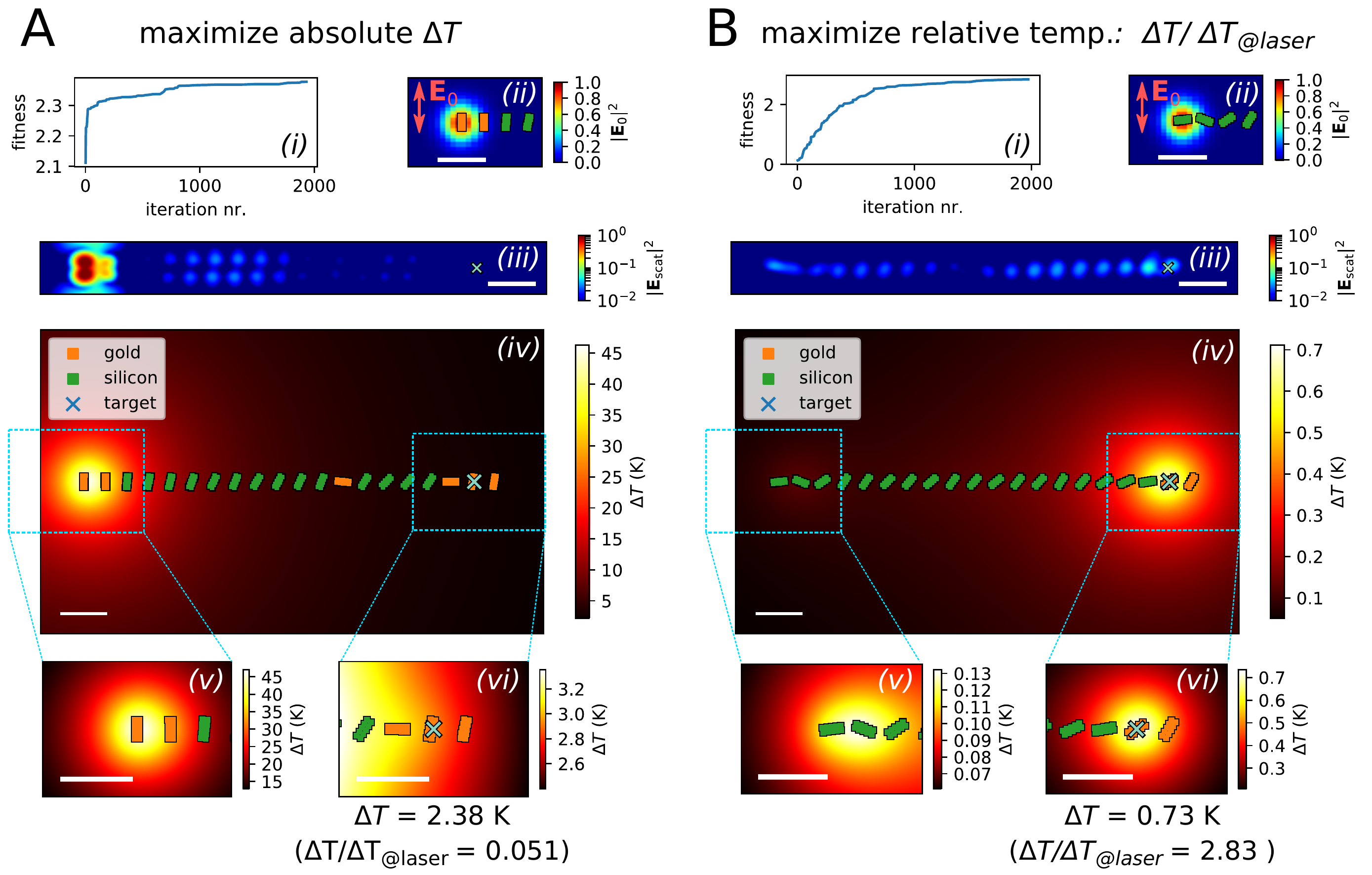}
\caption{(color online)
Results of the {\sf EO} of a chain for a maximization of (A) 
the {\it absolute} temperature increase at the target location 
and (B) the {\it relative} temperature increase at the target position, 
normalized to $\Delta T_{@laser}$ at the location of the focused illumination.
(i) convergence of the optimization.
(ii) intensity of the incident field $\mathbf{E}_0$, 
with focal spot of waist $w_0=300$ nm at $(0,0,0)$ nm, linearly polarized along $OY$ ($\lambda_0$ = 600nm).
(iii) scattered field intensity on a logarithmic color scale, 
calculated 50nm above the chain of nano--rods (normalized to $|\mathbf{E}_0|^2$).
(iv) mapping of the temperature increase along the chain.
(v) and (vi) zooms around the focal spot and the location of the $\Delta T$ probe.
All temperature mappings are calculated at a height of $z=300$ nm.
White scale bars are 500 nm.
The optimization target location is indicated by a cross--shaped marker.
}
\label{fig:chain_opt}
\end{figure}

This problem is inspired by works which have shown that chains of both, dielectric
and metallic nano-particles can effectively guide light along 
relatively large distances\cite{girard_nearfield_2004,bakker_resonant_2017}.
Here, the aim of the optimization is to find a chain of \(20\) nano-rods 
(each rod of size $70\times 175\times 140$ $nm^{3}$) 
which delivers the highest possible temperature increase at a specific location 
far from a focused illumination with a wavelength of $\lambda_0=600$ $nm$.
The focal spot with a beam waist of $w_0=300$ $nm$ is set at the origin, 
centered on the first nano-rod (see also subplots (ii) in Fig.~\ref{fig:chain_opt}).
The rods lie on a glass substrate ($n=1.5$, $\kappa=0.8$ $Wm^{-1}K^{-1}$), 
in water ($n=1.33$, $\kappa=0.6$ $Wm^{-1}K^{-1}$).
The temperature increase $\Delta T$ is to be maximized at $(x=4500, y=0, z=300)$ $[nm]$.
The free parameters for the optimization are each rod's rotation angle $\alpha_i$, 
each rod's material (either gold or silicon, refractive indices taken 
from Refs.~\cite{Johnson:1972,palik_silicon_1997}) 
and the distance $D$ between the nano-rods (which are equidistant along the chain).
The geometry of the problem is depicted in Fig.~\ref{fig:sketch_opt}B.
It is obvious, that a systematic evaluation of all possible 
solutions is impossible, considering the 41 free parameters.
For the {\sf EO} algorithm we use a population of 50 individuals, which we evolve for 2000 iterations.
On an ordinary office PC (AMD FX-8350 CPU) one run never took longer than 24 hours.
Figure~\ref{fig:chain_opt} shows the results of the optimizations, 
the fitness as a function of the iteration number is shown in the subplots (i).
In order to verify the convergence, we ran the optimizations several times with random initial parameters.
The different runs yielded similar results, hence we conclude that the optimizations converged close to the global optimum.
\subsubsection{Maximize absolute temperature increase}
In the first run, the absolute temperature difference $\Delta T$ is the optimization target. 
The results are shown in figure~\ref{fig:chain_opt}A.
The {\sf EO} algorithm chose a spacing $D$ of approximately 250nm.
It placed two gold rods in the beginning of the chain, 
which are aligned with the incident light's polarization in order to obtain 
the strongest response and, consequently, the highest dissipation inside the metal.
Following the first two gold blocks, a chain of silicon rods is guiding light towards 
the target location, where further gold rods dissipate the arriving light into heat 
(see the scattered near-field in subplot (iii) of figure~\ref{fig:chain_opt}).
The heat radiated from the two initial blocks locally rises the temperature by more 
than $45^{\circ}$. At $x=4.5$ $\mu m$, the temperature still is increased by more then 2 degrees.
The temperature rise is shown in subplots (iv)-(vi) of figure~\ref{fig:chain_opt}.
The heat generation from the terminal gold elements in the chain is only 
contributing weakly to the total temperature increase, adding about $0.25^{\circ}$.
\subsubsection{Maximize normalized temperature increase}
In a second run, the goal of the optimization is set to the \textit{relative} 
temperature increase $\Delta T/\Delta T_{@laser}$.
We aim to maximize the temperature at the target location $(4500,0,300)$ $[nm]$ 
normalized to the temperature at the position of the illumination beam at (0,0,300) $[nm]$.
The results of this simulation are shown in figure~\ref{fig:chain_opt}B.
The spacing \(D\) was set by the {\sf EO} again to approximately 250 nm, 
which seems to be ideal for guiding light trough the chain at the illumination wavelength $\lambda_0$ =600 nm.
However, in contrast to the maximization of the absolute temperature, 
the {\sf EO} algorithm chose silicon for the entire chain and placed two gold elements only at its very end.
The first two nano--rods were furthermore rotated perpendicular to the light's 
polarization in order to minimize their optical response and hence reduce dissipation in the early chain.
If the silicon-rod at (0,0,0) were oriented along the polarization 
of $\mathbf{E}_0$, it would induce a temperature increase of $\approx 1.4^{\circ}$ at $(0,0,300)$ $[nm]$.
Despite their horizontal orientation, the first two silicon rods 
effectively couple light into the chain, through which it is guided towards the gold-rods at its end.
There the light is dissipated inside the metal, which acts as a local heat source just below the target position.
The solution, found by the \textsf{EO}, has a very low $\Delta T_{\text{@laser}}$ $\approx 0.13^{\circ}$ 
at the origin, but increases the temperature at the target location by $\Delta T$ $\approx 0.73^{\circ}$.Compared to $\Delta T_{\text{\@laser}}$, this is \(\times 2.8\) higher.
\section{Conclusion}
To conclude, we have presented new 
geometries based on arrays of subwavelength metallic nanoparticles organized 
at the surface of a dielectric substrate that results in thermoplasmonic metasurfaces. 
The different kinds of arrays discussed here make possible an on-demand 
and spatially controlled rise of temperature by the mean of the incident wavelength and polarization. 
This approach, based on localized resonances, is complementary to high order plasmonic 
resonances in larger 2D cavities that allow for an all-optical and polarization dependent 
control of the temperature landscape in their vicinity \cite{Cuche:2017}. 
The formalism used in this work is optimized for the direct space applications 
and provides convenient analytical formula, with which the mechanisms for converting light energy 
into heat can be intuitively represented for both spherical and elongated metallic particles. 
Using anisotropic polarizabilities, our approach reveals the clear relationship 
between excitation parameters (laser power, polarization, and wavelength) and expected thermal effects 
(heat amount, temperature, ...). 
Interestingly, the extension of this formalism by means of a volume 
discretization procedure of the metallic structures placed in the vicinity of a solid--liquid interface, 
supplies a numerical test bench for future applications of complex metasurfaces 
in thermoplasmonics, such as those encountered in micro-fluidic environments for nano-biology.
Finally, we demonstrated that evolutionary optimization together with nano-optical 
simulations allows finding geometric assemblies for complex thermoplasmonic problems.
Using an appropriately formulated problem, a hybrid--material chain of silicon 
and gold nano-rods can be optimized such, that the optical energy, 
delivered by a focused illumination spot, is transferred towards a distant location where it is locally transformed into heat.

\newpage
{\bf Acknowledgments}: 
This work was supported by the Agence Nationale de la Recherche (ANR) (Grants ANR-13-BS10-0007-PlaCoRe),
the Programme Investissements d'Avenir under the program ANR-11-IDEX-0002-02, reference ANR-10-LABX-0037-NEXT,
and the computing center CALMIP in Toulouse.

\end{document}